\documentclass[12pt]{article}

\usepackage{amsmath,amssymb}
\usepackage{bm}
\usepackage{graphicx}
\usepackage{dcolumn}
\usepackage{bm}

\begin{document}
\baselineskip= 24pt
\begin{center}
{\large Tension Dynamics and Linear Viscoelastic Behavior \\
of a Single Semiflexible Polymer Chain} \\
\vskip 0.2 true cm
{T. Hiraiwa, and T. Ohta}
\vskip 0.2true cm
{Department of Physics, Graduate School of Science, Kyoto University, Sakyo-ku, Kyoto 606-8502, Japan}
\end{center}
\date{\today}


%

\begin{abstract}
We study the dynamical response of a
 single semiflexible polymer chain based on the theory developed by Hallatschek et al. for the wormlike-chain model. 
The linear viscoelastic response under oscillatory forces acting at the two chain ends is derived analytically as a function of the oscillation frequency $\omega$. 
We shall show that the real part $J'$ of the complex compliance $J=J'+iJ''$ in the low frequency limit 
 $\omega \to 0$ is consistent with the static result of Marko and Siggia whereas the imaginary part $J''$ exhibits  the power-law dependence $\omega^{+1/2}$. On the other hand, these compliances decrease
as $\omega^{-7/8}$ for the
high frequency limit $\omega \to \infty$.
These are different from those of the Rouse dynamics.
 A scaling argument is developed to understand these novel results.
\end{abstract}

\section{INTRODUCTION}
The recent experimental advances in the manipulation of single molecules,
such as optical tweezers and atomic force microscopy together
with single-molecule fluorescence \cite{Wang, Ladoux, Cocco, Ritort},
have enabled us to carry out mechanical and relaxational measurement 
in the nano-scale 
with piconewton sensitivity \cite{Strick}
in both
equilibrium and non-equilibrium conditions.
For example, 
static force-extension measurements of stretching of a single polymer chain have been 
carried out \cite{Marko,Wang,Bouchiat,Murayama}.
As a non-equilibrium dynamics, 
the viscoelastic properties or the elastic and dissipative properties 
have also been
studied \cite{Sakai,Khatri2,Kawakami1,Kawakami2,Kawakami3}.
Such experiments
have revealed more detailed properties of single molecules
that are difficult to obtain in bulk experiments due to
the average taken over molecules and time.
Therefore, these investigations lead to better understanding of 
the hierarchical structure of soft matter 
and the relationship between the molecular morphology and the functionality of biological molecules 
\cite{Ritort,Yamada}.

One of the characteristic features of soft matter such as
polymers or membranes is that they often have several length scales.
Even in a single polymer chain if the chain is semi-flexible, there are at least two length scales,
i.e., the persistence length and the total chain length.
In the several experiments of single polymer chains, 
the semiflexibility, i.e., the stiffness, is an important factor \cite{Marko,Sakai,Bouchiat}.
In fact, the wormlike-chain model, 
which is a model of a semiflexible polymer \cite{Kratky,Fixman,Marko}, 
explains many experimental results
considerably better than the flexible polymer chain model \cite{Edwards} 
particularly in the situation such as
the highly-stretching limit in the force-extension measurement
and the high wave-number limit of the dynamic structure factor, 
and so on \cite{Ladoux,Marko,Sakai,Bouchiat,Winkler1,Harnau}.
We emphasize that the rigidity effect can be enhanced in the above limits even for flexible polymers 
with a weak stiffness and that discrepancy appears between experiments 
and the theory based on a purely flexible model \cite{Harnau}. 
Therefore, investigation of the nonlinear dynamics due to the stiffness is necessary not only 
for semiflexible polymers but also for flexible polymers.

Despite the above fact as well as their fundamental interest 
in the field of mesoscopic physics
and their importance to the material and biological application,
semiflexible polymer chains have not been studied intensively 
especially for the dynamics because of the strong nonlinearity contained in the wormlike-chain model. 
Most of the theoretical studies of single polymers have been made 
in the limiting cases of either very flexible polymers or rigid rods \cite{DoiEd}.
So far, computer simulations have been carried out 
for a stiff chain or a semiflexible chain \cite{Somasi, Yoshinaga, Chatt, Morrison}. 

Static theories of a semiflexible polymer chain are summarized as follows.
Marko and Siggia derived the static force-extension relation 
based on the wormlike-chain model \cite{Marko}.
Other statistical properties, such as the distribution function of the end-to-end distance,
have also been investigated \cite{Wilhelm,Chirikjian,Hamprecht}.
Improvement of the wormlike-chain models has been proposed to
examine the static properties \cite{Bouchiat,Winkler4}.

On the other hand, as mentioned above, analytical approaches to non-equilibrium 
dynamics of a semiflexible 
single polymer chain are limited.
Some of the previous works
have employed an approximation of linearization 
for the inextensibility constraint \cite{Saito}.
This linearization
neglects non-uniformity of the line tension along the chain and has been
applied to stretched polymers \cite{Winkler1,Bohbot-Raviv,Winkler2,Winkler3}.

Recently, Hallatschek et al. \cite{Hall1, Hall2} 
have formulated the force-extension
theory for the wormlike-chain dynamics without linearization of
the inextensibility condition
introducing the concept of tension propagation.
They consider a weakly bend situation 
and use a kind of multi-scale perturbation methods.
The theory has been applied to the relaxation of an elongated chain after removing
an external force \cite{Ober1, Ober2}.

Finally, it is also mentioned that,
as a previous theoretical method,
 the scaling approach to a semiflexible polymer chain \cite{Everaers,Hall2,Ober2}, 
which was successful for  flexible chains \cite{deGennes-text,Pincus,Khatri}.

In the present paper, we develop the linear viscoelastic theory of a strongly 
pre-stretched single semiflexible polymer chain. 
We consider the situation such that an oscillatory force in addition to a constant force is applied to the two end of a wormlike-chain.
Based on the method by Hallatschek et al. we derive the analytic representation of
the complex compliance and the complex modulus.
It will be shown that 
the frequency dependence is quite different from that of the Rouse model \cite{Khatri}.
The preliminary results have been published in Ref. \cite{Hiraiwa}. 
We apply a scaling analysis to understand the physical insight of the results.

The outline of the paper is as follows:
In Section \ref{sec:model},
we present the dynamical model of the wormlike-chain
and the tension-propagation equation is derived based on the method by 
Hallatschek et al. \cite{Hall1, Hall2}.
In Section \ref{sec:result},
the complex compliance and the complex modulus are obtained analytically.
In Section \ref{sec:Rouse}, 
the compliance and the modulus in the Rouse dynamics are given for comparison.
In Section \ref{sec:scaling}, 
the scaling approach is applied
to both the weak-bending wormlike-chain dynamics and the Rouse dynamics.
Summary and discussion are given in Section \ref{sec:con}.

\section{WORMLIKE-CHAIN MODEL AND THE RESPONSE TO THE OSCILLATORY FORCE} \label{sec:model}
\subsection{Dynamics of the wormlike-chain model}
The effective Hamiltonian for the wormlike-chain is given by \cite{Kratky}
\begin{equation}
H_{WLC}=\frac{\kappa}{2}\int^{L}_{0}ds \left | \frac{d^2 {\bm r}}{ds^2} \right | ^2,
 \label{eq:H}
\end{equation}
with the constraint
\begin{equation}
 | {\bm r'}(s,t) |^2 =1,
 \label{eq:cons}
\end{equation}
where $t$ denotes the time, $s$ is the length along the chain from one end, 
$L$ is the total length and ${\bm r}(s,t)$  represents the conformation of the chain. 
The positive constant $\kappa$ is the bending rigidity. 
The prime indicates the derivative with respect to $s$.
The constraint (\ref{eq:cons}) can be incorporated into the Hamiltonian as
\begin{equation} \label{eq:modelHam}
 H_{WLC}=\frac{\kappa}{2}\int^{L}_{0}ds \left | \frac{d^2 {\bm r}}{ds^2} 
\right | ^2 + \frac{1}{2} \int_{0}^{L}ds f(s,t) \left|\frac{d{\bm 
r}}{ds}\right|^2 \ ,
\end{equation}
where  $f(s,t)$ is the Lagrange multiplier for the constraint (\ref{eq:cons}) and 
is interpreted as the line-tension. 
By assuming the over-damped motion, the stochastic equation of motion of a chain is given by
\begin{equation} \label{eq:dynamics1}
 \zeta \partial _{t} {\bm r} (s,t) = - \kappa {\bm r''''} + ( f (s,t) {\bm r'}(s,t) )' 
 + {\bm g}(s,t) + {\bm \xi(s,t)} ,
\end{equation}
where  the friction coefficient $\zeta$ is a $3\times3$ matrix with the 
components $\zeta_{ij}$ ($i,j=x,y,z$) and 
${\bm g}(s,t)$ represents the external force.
The random force ${\bm \xi(s,t)} $ obeys the Gaussian white statistics:
\begin{gather}
< \xi _{i} (s,t) > = 0, \\
< \xi_{i}(s,t) \xi_{j}(s',t') > = 2 k_{B}T \zeta_{ij} \delta (s-s') \delta (t-t') 
\end{gather}
with $k_B$ the Boltzmann coefficient and $T$ the absolute temperature.
The equation of motion (\ref{eq:dynamics1})
is the same as that employed by Liverpool \cite{Liverpool}.

A remark is now in order. A stiff filament with an internal friction has been studied where 
the friction is supposed to arise from the internal conformation rearrangement of 
the filament with a finite radius \cite{Marko02}. It is emphasized here that we have not 
introduced such an additional friction in eq \ref{eq:dynamics1}. As described below, 
the constraint eq \ref{eq:cons} produces a strong nonlinear coupling between 
the longitudinal (parallel to the external force) and the transverse components of the conformation, 
which causes an 
energy dissipation whose magnitude  is comparable with the typical elastic energy.

\subsection{Weak bending approximation and multiple scale analysis}
Now we follow the theory developed  by Hallatschek, Frey and Kroy \cite{Hall1, Hall2}. 
They consider the situation 
such that the chain is elongated by the force $f$ applied to the ends. 
The smallness parameter is introduced as $\epsilon\equiv k_BT/(\kappa f)^{1/2}$. 
The conformation vector ${\bm r}(s, t)$ is divided into two components. 
One is parallel to the elongation direction 
(along the x-axis) and the other is perpendicular to it, i.e., ${\bm r}(s, t)=(s-r_{\parallel}, {\bm r}_{\perp})$. 
The basic approximation is the weak bending approximation such that ${\bm r}_{\perp}'(s,t)^2 = O(\epsilon ) \ll 1$. 
In this situation we have $r'_{\parallel}=(1/2) ({\bm r}'_{\perp})^2 + O(\epsilon^2)$. 
Hallatschek et al. \cite{Hall1, Hall2} have introduced a concept of stored excess length defined by
\begin{align}
\rho(s, t) = \frac{1}{2}({\bm r}'_{\perp})^2. \label{eq:sel}  
\end{align}
Since the parallel component of the end-to-end distance is given by 
$R_{\parallel}\equiv L-(r_{\parallel}(L)-r_{\parallel}(0))$, we obtain the relation 
\begin{equation}
< \Delta R_{\parallel}>(t) = -\int^{L}_{0} <\Delta \rho> (s,t) ds + o(\epsilon)  ,
\end{equation}
where $\Delta R_{\parallel}$ and $\Delta \rho$ indicate the deviation from some reference state 
and $<..>$ means a statistical average. 

The Langevin equation (\ref{eq:dynamics1}) is split into two equations for $r_{\parallel}(s, t)$ 
and ${\bm r}_{\perp}(s, t)$ 
with the scalar friction coefficients $\zeta_{\parallel}$ and $\zeta_{\perp}$ respectively.
The equation of the transverse motion is given by
\begin{equation} \label{eq:perpmotion}
 \zeta _{\perp} \partial _t {\bm r}_{\perp} =
  - \kappa {\bm r''''}_{\perp} +(f(s,t){\bm r'}_{\perp})' 
       +{\bm g}_{\perp}+{\bm \xi _{\perp}} \ ,
\end{equation}
where the external force ${\bm g}$ and the random force ${\bm \xi}$ are 
divided into the longitudinal and transverse components
as ${\bm g}(s,t) = (g_{\parallel},{\bm g}_{\perp})$ and 
${\bm \xi}(s,t) = (\xi_{\parallel},{\bm \xi}_{\perp})$, respectively.
Taking the first derivative with respect to $s$ for the both sides of eq \ref{eq:dynamics1}, 
the equation of the longitudinal motion is given by
\begin{align} \label{eq:paramotion}
 \zeta_{\parallel}&\partial _{t} r_{\parallel}' = +(\zeta _{\parallel} - \zeta _{\perp}) 
   ({\bm r'}_{\perp}\cdot \partial _t {\bm r}_{\perp})' \notag \\ 
  &- \kappa r_{\parallel}''''' - f''(s,t) + (f(s,t)r'_{\parallel})'' 
  - g_{\parallel}' - \xi_{\parallel}' \ .
\end{align}
Note that the sign in front of $\xi_{\parallel}' $ is minus because of the relation ${\bm r}=(s-r_{\parallel},{\bm r}_{\perp})$.
In these expressions,
$o(\epsilon^{1/2})$ terms and $o(\epsilon^{1})$ terms 
are neglected in eq \ref{eq:perpmotion} and eq \ref{eq:paramotion}, respectively. 
This set of equations is solved by a perturbation expansion together with the multiple scale analysis by introducing 
two scaled variables 
$s_{\text{s}}=s$ and $s_{\ell}=\epsilon^{1/2}s$.
Noting that the ratio of the relaxation rate of $r_{\parallel}$ to that of ${\bm r}_{\perp}$ 
is $O(\epsilon^{-1/2})$, one may apply an adiabatic approximation for $r_{\parallel}$. 
Furthermore, the local equilibrium approximation is employed such that the degrees of freedom 
in the length scale $s_{\text{s}}$
is relaxed for a given constraint 
for the larger scale $s_{\ell}$.
In this way, one obtains the following set of equations 
\begin{align}
- \frac{1}{k_BT} <\Delta \bar{\rho}(s,t)> =  \int^{\infty}_{0} \frac{dq}{\pi} \left \{\frac{1-\exp( -A(q, s, t))}{\kappa q^2+f_0} \right.  \notag\\
    \ \left. - \frac{2 q^2}{\zeta_{\perp}} \int_{0}^{t} d \tilde{t}  
   \exp \left(-A(q, s, t)+A(q, s, \tilde{t}) \right) \right \} 
     \label{eq:nonlinear1}  
\end{align}
and
\begin{equation}    
 \label{eq:nonlinear2} 
<\Delta \bar{\rho}>(s, t)= -\frac{1}{\zeta_{\parallel}} \partial _s^2 
F(s,t) \ ,
\end{equation}
where $q$ is the wave number representing modulations of 
the conformation ${\bm r}_{\perp}(s, t)$ and
 \begin{eqnarray} \label{eq:defF}
F(s, t)=\int_0^t d \tilde{t} f(s,\tilde{t}),
  \end{eqnarray}
 \begin{eqnarray} \label{eq:defA}
A(q, s, t)=2q^2 \Bigg( \kappa q^2 t + F(s,t)\Bigg) /\zeta_{\perp} \ .
  \end{eqnarray}
The quantity  $<\Delta \bar{\rho}>(s, t)$ is the bulk value of  $<\Delta \rho>(s, t)$. See 
Ref. \cite{Hall1} 
for details. 
We consider the situation such that the polymer chain is in a steady condition under 
 a constant force $f_0$ applied at the ends till $t=0$ and then another time dependent force $\Delta f (s,t)$
is switched on at $t=0$,
i.e., $f(s,t)=f_0+\Delta f(s,t)$ for $t>0$.

The tangential vector at the chain ends is approximated to be parallel to the direction of the external force. 
This is justified in the weak bend limit \cite{Hall1}.
The time-integral of the force along the polymer chain is given by 
\begin{align}
 F(s,t) = F_0(t)+\Delta F(s,t) ,
 \label{eq:Fst}
\end{align}
where $F_0(t) \equiv f_0 t$ and
\begin{align}
\Delta F(s,t) \equiv \int^{t}_{0} d \tilde{t} \Delta f (s,\tilde{t}) .
\end{align}

\subsection{Characteristic length and time}
By comparing three terms in  (\ref{eq:modelHam}), one notes that there are three characteristic lengths
\begin{align} \label{eq:lp}
 &\ell_p=\frac{\kappa}{k_BT} \\
 &\ell_f=\frac{k_BT}{f} \label{eq:lf} \\
 &\xi = \left( \frac{\kappa}{f}\right)^{1/2} \ ,  \label{eq:xi}
\end{align}
where $\ell_p$ is the persistence length of the chain and 
$\xi$ has a meaning of the ``screening'' length. 
In a linear response as we study in the present paper, 
the constant force $f_0$ should be used for $f$.
The total length of the chain $L$ is also a characteristic length.
The smallness parameter of the weak bending limit $\epsilon$ can be 
rewritten as follows
\begin{equation} \label{eq:epsreform}
 \epsilon = \frac{\xi}{\ell_p}=\frac{\ell_f}{\xi}=\left(\frac{k_BT}{\ell_p f}\right)^{1/2} \ .
\end{equation}
This indicates that the magnitude of the characteristic lengths has a definite order for 
$\epsilon \ll 1$ as
\begin{equation}
 \ell_f \ll \xi \ll \ell_p 
 \lesssim L \ .
\end{equation}
Hereafter we ignore the shortest one $\ell _f$.

Comparing each term in the Langevin equation (\ref{eq:dynamics1}), we obtain the following characteristic times 
\begin{align} \label{eq:tau1}
 &\tau_1 = \frac{\ell^4 \zeta_{\perp}}{\kappa} \\
 &\tau_2 = \frac{\ell^2 \zeta_{\perp}}{f} \label{eq:tau2}
\end{align}
with $\ell$ a length scale.
Substituting $\ell=\xi$ into eq \ref{eq:tau2}, we obtain
\begin{equation} \label{eq:tauN}
 \tau_{\xi} = \frac{\kappa \zeta_{\perp}}{f^2} \ .
\end{equation}
Substituting $\ell=\ell_p$ into eq \ref{eq:tau1}, we have
\begin{equation} \label{eq:tau3}
 \tau_p=\frac{\ell_p^3 \zeta_{\perp}}{k_BT}=\frac{\kappa^3 \zeta_{\perp}}{(k_BT)^4} \ .
\end{equation}
Note that this is the only characteristic time which does not contain neither $f$ nor $L$.

\subsection{Tension dynamics}
In this subsection, we focus on the propagation of the line tension $f(s,t)$ or $F(s,t)$.
Here it is mentioned that this concept itself can also be applied to a 
flexible polymer chain \cite{Sakaue}.
Combining eqs \ref{eq:nonlinear1} and \ref{eq:nonlinear2}, 
the tension propagation equation is obtained as the closed form with respect to $F$;
\begin{align}
 \frac{\pi}{\zeta_{\parallel} k_BT} \partial_s^2 F(s,t) =  \int^{\infty}_{0} dq \left \{\frac{1-\exp( -A(q, s, t))}{\kappa q^2+f_0} \right.  \notag\\
    \ \left. -\frac{2 q^2}{\zeta_{\perp}} \int_{0}^{t} d \tilde{t}  \exp \left(-A(q, s, t)+A(q, s, \tilde{t}) \right) \right \}  \ .
     \label{eq:nonlinear3}  
\end{align}
This equation is 
rewritten in terms of the dimensionless quantities as 
\begin{gather}
 K \partial_{\hat{s}}^2 \hat{F}(\hat{s},\hat{t}) 
= \int^{\infty}_{0} d\hat{q} \left\{ \frac{1-\exp( -\hat{A}(\hat{q}, \hat{s}, 
 \hat{t}))}{\hat{q}^2+1} \right. \notag \\
  \ \left. -2 \hat{q}^2 \int_{0}^{\hat{t}} d \tilde{t}  
\exp \left(-\hat{A}(\hat{q},\hat{s},\hat{t})+\hat{A}(\hat{q},\hat{s},\tilde{t})\right) 
 \right\} \ ,  \label{eq:nonlinear5}
\end{gather}
where
\begin{gather} \label{eq:qsscale}
 \hat{q} = \xi q \ , \\
 \hat{s} =\epsilon^{1/2} s \xi^{-1} \ , \label{eq:eliofeps} \\
 \hat{t} = t/ \tau_{\xi} \ .
\end{gather}
$K = \pi/\hat{\zeta}$ is just a numerical factor with $\hat{\zeta} \equiv \zeta_{\parallel}/\zeta_{\perp}$.
The total length $L$ is now rescaled as $\hat{L} =\epsilon^{1/2}L \xi^{-1}$.
The scaled functions $\hat{A}$ and $\hat{F}$ are given by
\begin{equation}
 \hat{A}(\hat{q}, \hat{s}, \hat{t})=2 \hat{q}^2 
\Bigg(\hat{q}^2 \hat{t} + \hat{F}(\hat{s},\hat{t})\Bigg)
\end{equation}
and
\begin{equation} \label{eq:FhatFtrans} 
\hat{F}(\hat{s}, \hat{t}) =\frac{\xi^2}{\kappa \tau_{\xi}} F(s,t) 
 =  \int_0^{\hat{t}} d \tilde{t} 
\frac{f(\xi \hat{s},\tau_{\xi} \tilde{t})}{f_0}  \ . 
\end{equation}

We assume that $\Delta f (s,t)$ is sufficiently small and apply the linearization approximation to 
(\ref{eq:nonlinear5}). That is, we substitute (\ref{eq:Fst}) into
(\ref{eq:nonlinear5}) and retain the terms 
up to the first order with respect to $\Delta F$ so that we obtain
\begin{eqnarray} 
 K \partial_{\hat{s}}^2 \Delta \hat{F} + \int^{\hat{t}}_{0}d 
  \tilde{t} \Delta \hat{F}(\hat{s},\hat{t}-\tilde{t}) M(\tilde{t}) = 0 ,
 \label{eq:linear}
\end{eqnarray}
where the memory function $M(\hat{t})$ is given by
\begin{align}
 M(\hat{t}) & \equiv 4 \int^{\infty}_{0}dq \left\{ q^4 e^{-2q^2(q^2+1)\hat{t}} - \frac{q^2}{q^2+1} \delta(\hat{t}) \right\} .
\end{align}
The asymptotic behavior is given by $M(\hat{t}) \sim \hat{t}^{-\beta}$
 with $\beta=5/4$ for $\hat{t} \to 0$ and $\beta=5/2$ for $\hat{t} \to \infty$. 

Equation (\ref{eq:linear}) is to be solved under the boundary conditions specified  by $\Delta F(0,t)$
and $\Delta F (L,t)$. In what follows, we consider the symmetric case 
that $\Delta F(0,t)=\Delta F (L,t) \equiv \Delta F (t)$.
Applying the Laplace transformation with respect to $t$ to 
eq \ref{eq:linear}, we obtain
\begin{equation} \label{eq:raplace}
 K \partial _{\hat{s}}^2 \Delta \tilde{F}(\hat{s},z)  + N(z) \Delta 
  \tilde{F}(\hat{s},z)  = 0 \ ,
\end{equation}
where $\tilde{F}(\hat{s},z)$ denotes the Laplace transform of $\hat{F}(\hat{s},t)$ and 
\begin{equation} 
 N(z) \equiv 4 \int_{0}^{\infty} dq \left\{ \frac{q^4}{2q^2(q^2+1)+z} 
      -\frac{1}{2} \frac{q^2}{q^2+1} \right\}  \label{eq:Nbar}  
\end{equation}
is the Laplace transform of $M(t)$.
The asymptotic form of $N(z)$ is given as follows.
For $\omega \to \infty$, from eq \ref{eq:Nbar} we obtain the 
following equations after some manipulation
\begin{align} 
 \rm{Re} N(\pm i\omega \tau_{\xi})  &=- 2 S_1(\omega \tau_{\xi})^{1/4} , \notag \\
 \rm{Im} N(\pm i\omega \tau_{\xi})  &= \mp 4 S_2(\omega \tau_{\xi})^{1/4}    \label{eq:NbarL}  
\end{align}
with
$S_1=\int^{\infty}_0dq(4q^8+1)^{-1}\approx 0.863$ and 
$S_2=\int^{\infty}_0dq q^4(4q^8+1)^{-1} \approx 0.179$.
It is readily shown that the Taylor expansion of N(z)
 with respect to $z$ breaks down and therefore N(z) is not analytic at $z=0$. 
The correct expansion
is obtained after some manipulation as follows 
\begin{equation} \label{eq:N0limit}
 N(\pm i\omega \tau_{\xi}) = - 2 S_3(\omega \tau_{\xi})^{3/2} 
  \mp i S_4 (\omega \tau_{\xi})^1 \ ,
\end{equation}
where $S_3 = \pi/8$ and $S_4 = \pi/4$.

We consider the case that the force $\Delta f(t)$ at the boundaries is 
oscillatory as
$\Delta f(t) =f_A \sin(\omega t)$ with the amplitude $f_A$ and the frequency $\omega$.
The scaled form of $\Delta F$ at the boundaries is given by
\begin{equation}
 \Delta \hat{F}(\hat{t}) = (\omega \tau_{\xi})^{-1}\frac{f_A}{f_0}
 \left[ 1-\cos \left( \omega \tau_{\xi} \right)\right]
\end{equation}
and the Laplace transform is 
\begin{equation}
 \Delta \tilde{F}(z)  = \frac{f_A}{f_0}
 \frac{ (\omega \tau_{\xi})}{z [z^2 + (\omega \tau_{\xi}) ^{2}]}.
\end{equation}
The solution of eq \ref{eq:raplace} can be represented as
\begin{equation}
 \Delta \tilde{F}(\hat{s},z)  = \Delta \tilde{F}(z)  \times 
\frac{\cos\left(B(z)(2\hat{s}-\hat{L})\right)}{\cos\left( B(z) \hat{L}\right)},
  \label{eq:Fbar}  
\end{equation}
where $B(z)= (N(z)/4K)^{1/2}$.
One needs to evaluate the inverse Laplace transform of eq \ref{eq:Fbar}
\begin{equation} \label{eq:time-space}
 \Delta \hat{F}(\hat{s},t) = \frac{1}{2 \pi i} \int_{c- i \infty}^{c+i \infty}\Delta \tilde{F}(z) 
 \frac{\cos\left(B(z)(2\hat{s}-\hat{L})\right)}{\cos\left( B(z) \hat{L}\right)} e^{z t} dz .
\end{equation}
This will be carried out in the next section.

\section{ANALYTICAL RESULTS} \label{sec:result}

Now, we study the response of the end-to-end distance to the oscillatory force.
The average end-to-end distance $\Delta R(t)$ which is a deviation 
from that of the steady state under the constant force $f_0$ is given by
\begin{align}
\Delta R (t) &= -\int ^{L}_{0} ds <\Delta \bar{\rho}>(s, t) 
 \label{eq:etedef} \\
&=\frac{1}{\zeta_{\parallel}}\{\partial_s F(s,t)|_{s=L} - 
\partial_sF(s,t)|_{s=0}\} \ .
\label{eq:DelR}
\end{align}
Hereafter, for abbreviation, we represent
the statistical average of the 
end-to-end distance as $\Delta R(t)$ 
without the brackets $<\cdot>$, the bar $\bar{\cdot}$ and the parallel mark $\cdot_{\parallel}$.
Substituting the solution (\ref{eq:time-space}) into eq \ref{eq:DelR} 
together with eq \ref{eq:FhatFtrans}, 
we obtain the time evolution of the average end-to-end distance under 
the given boundary condition.

Since we are concerned with the asymptotic behavior $t \rightarrow +\infty$, 
we consider only the poles on the imaginary axis $z = 0, \pm i \omega \tau_{\xi}$ 
to carry out the inverse Laplace transform. 
The final result can be written as
\begin{equation}
\frac{\Delta R(t)}{L}=\frac{f_A}{f_0}
\left[ \hat{J}'(\omega)\sin(\omega t) - \hat{J}''(\omega) \cos(\omega t) 
\right] \ .
\end{equation}
The scaled complex compliance 
is given by
\begin{eqnarray}
 \hat{J}'(\omega) = -\frac{2D}{\omega \tau} \rm{Im}\Bigg(\bar{B}(i\omega\tau_{\xi}) \tan(\bar{B}(i\omega\tau_{\xi}) )\Bigg) ,
                      \label{eq:J} 
\end{eqnarray}
\begin{eqnarray}
 \hat{J}''(\omega) =- \frac{2D}{\omega \tau}\rm{Re} \Bigg(\bar{B}(i\omega\tau_{\xi}) \tan( \bar{B}(i\omega\tau_{\xi}) )\Bigg),
                    \label{eq:JJ} 
\end{eqnarray}
where $\bar{B}(z) = \alpha N(z)^{1/2}/2$
with complementary dimensionless constants 
\begin{equation} \label{eq:alphadef}
\alpha \equiv 
\frac{\hat{\zeta}^{1/2}(k_BT)^{1/2}f_0^{1/4}L}{\pi^{1/2} {\kappa}^{3/4}} 
= \sqrt{\frac{\hat{\zeta}}{\pi \epsilon}} \frac{L}{\ell_p}
\end{equation}
and 
\begin{equation}
D \equiv  \frac{1}{2\pi^2} \frac{k_BT}{\sqrt{f_0\kappa}} = \frac{1}{2\pi^2} \epsilon \ .
\end{equation} 
The scaled elastic modulus $\hat{G}'$ and the scaled
loss modulus $\hat{G}''$ are obtained from $\hat{J}'$ and $\hat{J}''$ as follows
\begin{equation}
 \hat{G}'(\omega)=\frac{\hat{J}'(\omega)}{\hat{J}'(\omega)^2 + 
 \hat{J}''(\omega)^2} \ ,
\end{equation}
\begin{equation}
 \hat{G}''(\omega)=\frac{\hat{J}''(\omega)}{\hat{J}'(\omega)^2 + 
 \hat{J}''(\omega)^2} \ .
\end{equation}

\begin{figure}[h]
 \centering
 \includegraphics[width=7.1cm]{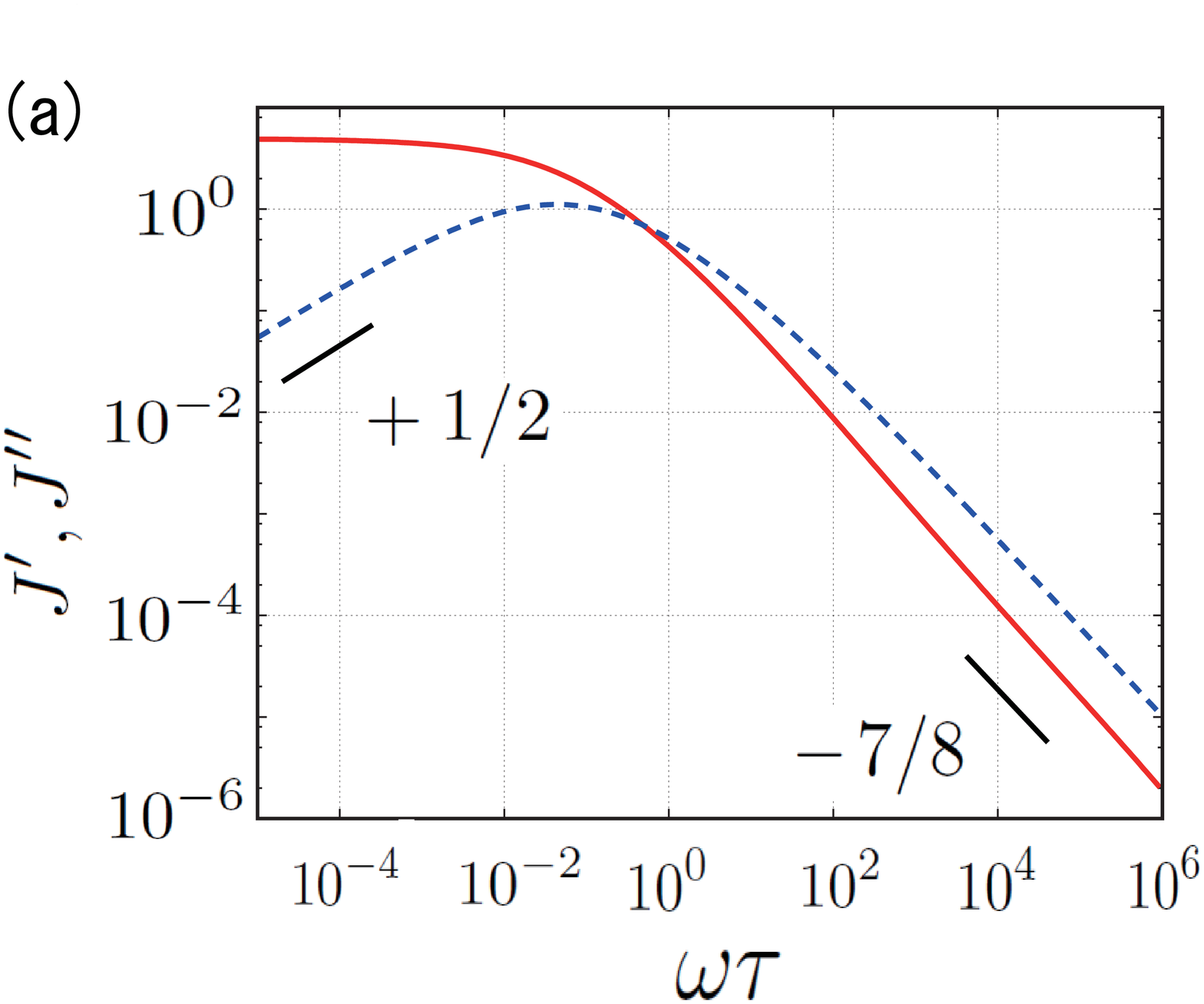}
 \includegraphics[width=7.1cm]{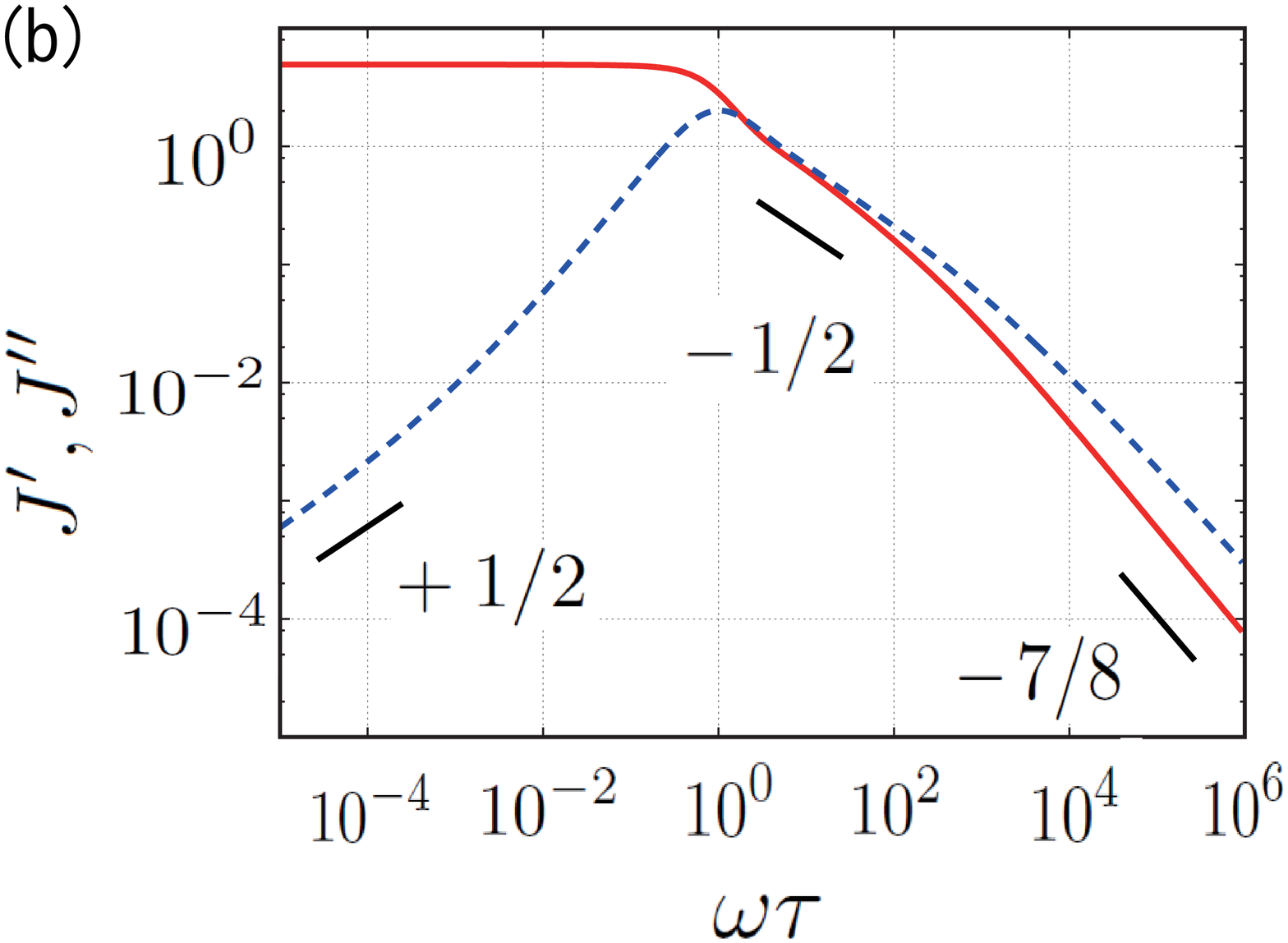}
  \caption{$\hat{J}'$ and $\hat{J}''$ as a function of $\omega \tau$ for $D=1$ and  (a) $\alpha=1.0$  and
 (b) $\alpha=100.0$. The full curve represents $\hat{J}'$ whereas the broken curve represents $\hat{J}''$.
 The characteristic time $\tau$ is defined by eq \ref{eq:tau}.   
   } 
 \label{graph:Js}
\end{figure}

\begin{figure}[h]
 \centering
 \includegraphics[width=7.5cm]{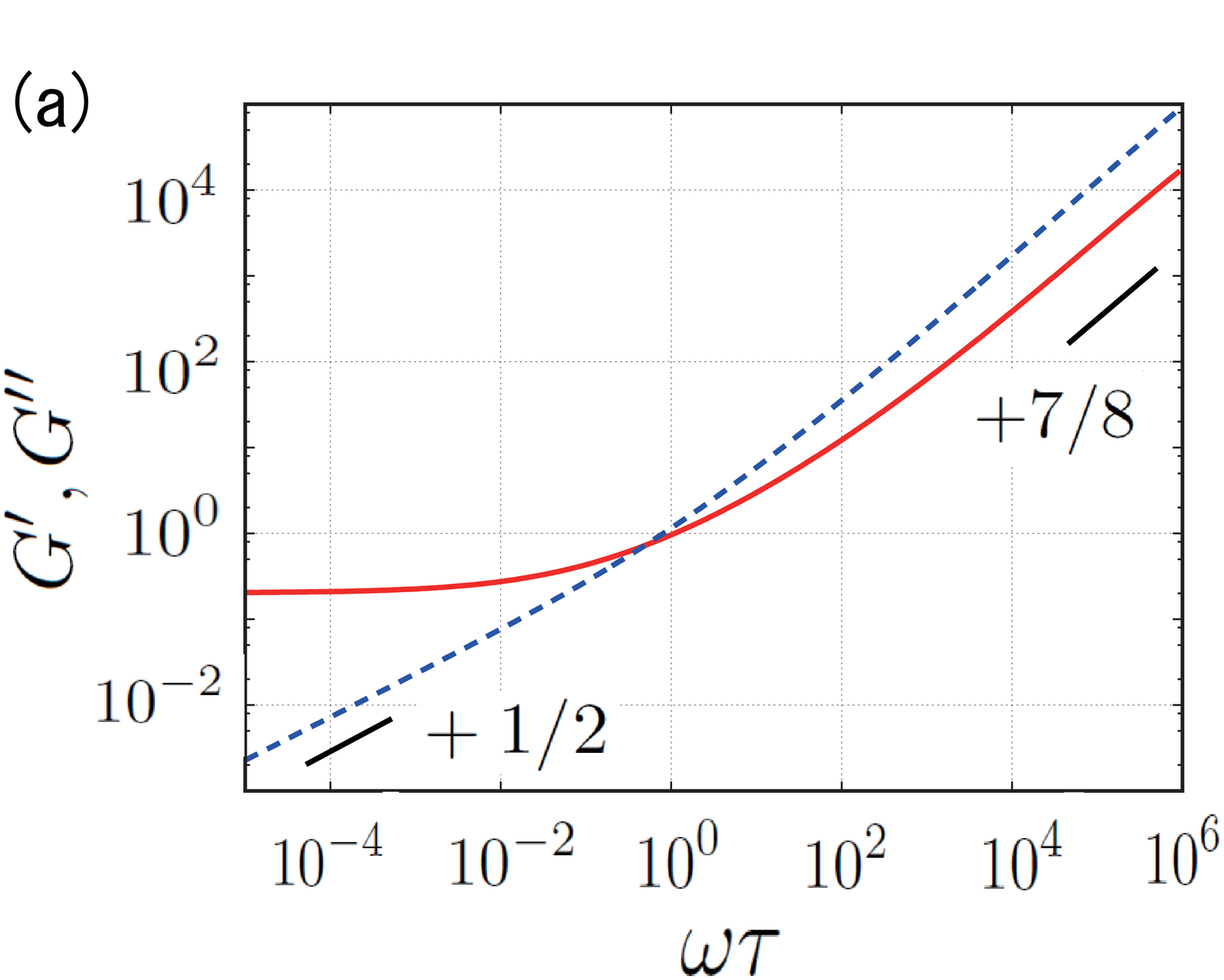}
 \includegraphics[width=7.5cm]{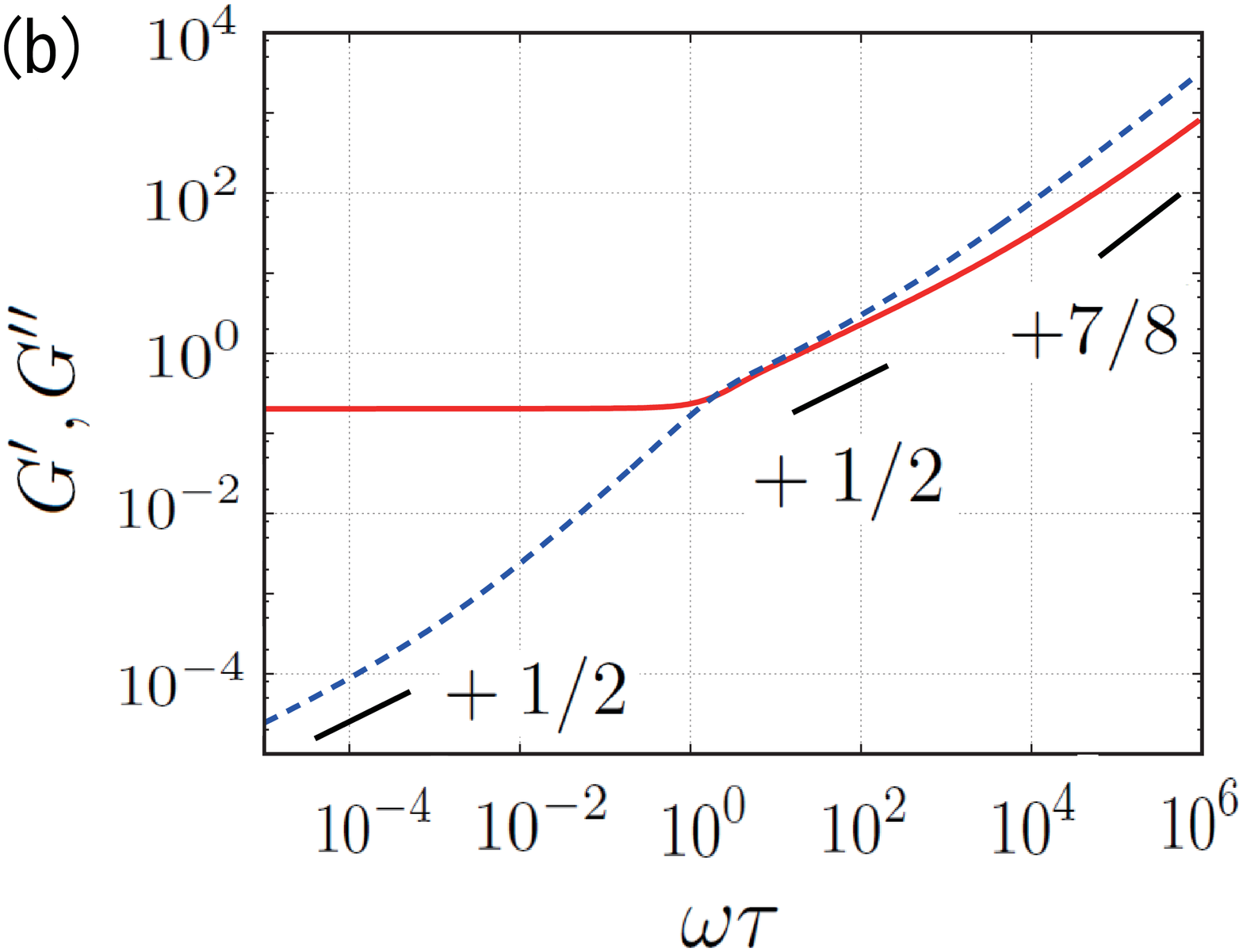}
 \caption{$\hat{G}'$ and $\hat{G}''$ as a function of $\omega \tau$ for $D=1$ and  (a) $\alpha=1.0$ and (b) 
 $\alpha=100.0$. The full curve represents $\hat{G}'$ whereas the broken curve represents $\hat{G}''$.
} 
 \label{graph:Gs}
\end{figure}

In the following, we introduce another characteristic time.
The linearized eq \ref{eq:linear} reduces to the following simple 
diffusion equation by employing the Markov 
approximation;
\begin{equation}
 K \partial_{\hat{s}}^2 \Delta \hat{F}(\hat{s},\hat{t}) - \frac{\pi}{4} 
  \partial_{\hat{t}} \hat{F} =0 \ .
\end{equation}
This implies that we may define
a new relaxation time by the following form as the time scale of the slowest mode,
just as the Rouse time in the continuous Rouse dynamics;
\begin{equation}  \label{eq:tau}
 \tau \equiv \frac{k_BT \zeta_{\parallel} L^2 }{4\pi^2 {\kappa}^{1/2} f_0^{3/2}}
= \frac{\alpha^2}{4 \pi}\tau_{\xi} \ .
\end{equation}
By this relation, 
the three parameters $\tau$, $\tau_{\xi}$ and $\alpha$ are not independent of each other.
In Figures \ref{graph:Js} and 
\ref{graph:Gs},
we choose $\alpha$ and $\tau$ as the independent parameters.

We examine the limiting behavior of $\hat{J}'$ and $\hat{J}''$. 
For the high frequency limit, substituting 
eq \ref{eq:NbarL} into eqs \ref{eq:J} and \ref{eq:JJ} 
and after some manipulation, we obtain
\begin{align}
\hat{J}'(\omega) &\sim \frac{4 S_1^{1/2} b (k_BT)^{1/2} f_0}{\pi ^{1/2} 
 \hat{\zeta}^{1/2} \kappa^{5/8} \zeta_{\perp}^{7/8} L} \omega^{-7/8} 
 \propto \kappa^{-5/8} (k_BT)^{+1/2} \omega^{-7/8} \ , \notag \\
\hat{J}''(\omega) &\sim\frac{4 S_1^{1/2} a (k_BT)^{1/2} f_0}{\pi ^{1/2} 
 \hat{\zeta}^{1/2} \kappa^{5/8} \zeta_{\perp}^{7/8} L} \omega ^{- 7/8} 
 \propto \kappa^{-5/8} (k_BT)^{+1/2} \omega^{-7/8} \label{eq:winftyexp}
\end{align}
and
\begin{equation}
\hat{J}'(\omega)/\hat{J}''(\omega) = b/a \approx 0.199 \ ,
\end{equation}
where $a \sim 0.721$ and $b \sim 0.142$ are the positive solutions of $(a+bi)^2 =1/2 + iS_2/S_1$. 
It should be noted that the unscaled complex compliance 
$J=L\hat{J}/f_0$ depends on neither $L$ nor $f_0$.

For the low frequency limit, substituting 
eq \ref{eq:N0limit} into eqs \ref{eq:J} and \ref{eq:JJ} 
and after some manipulation, we obtain
\begin{align} 
 \label{eq:NbarS1}  
 \hat{J}'(\omega) &\sim \frac{1}{4}\frac{k_BT}{\sqrt{\kappa f_0}} ,\\
 \hat{J}''(\omega)&\sim \frac{k_BT \zeta_{\perp}^{1/2}}{4f_0^{3/2}}\omega^{+1/2} 
 \propto \kappa^0 (k_BT)^{+1} \omega^{+1/2} \ .  \label{eq:NbarS2}  
\end{align}
Note that (\ref{eq:NbarS1}) is consistent with the result of Marko and 
Siggia for the static stress-strain relation \cite{Marko}, which is 
given by
\begin{equation}
 \frac{R(f_0)}{L}=1-\frac{1}{2} \frac{k_BT}{\sqrt{\kappa f_0}} \ .
\end{equation}
From this, we have
\begin{equation}
 \frac{R(f_0 (1+ \delta))}{L} -\frac{R(f_0)}{L} 
 = \frac{\delta}{4} \frac{k_BT}{\sqrt{\kappa f_0}} +O(\delta^2) .
\end{equation}

Equations (\ref{eq:J}) and (\ref{eq:JJ}) give us the complex compliance as a function of $\omega$.
Figures \ref{graph:Js}(a) and 
\ref{graph:Js}(b) show the compliances $\hat{J}'$ and $\hat{J}''$ for  $\alpha=1$ and  $\alpha=100$, respectively.
As mentioned above, the compliances  exhibit the fractional  power law behavior for the high frequency
 and $\hat{J}'$ is consistent with the static result of the wormlike-chain for $\omega \to 0$. 
The difference for the simple Maxwell-like elasticity is more evident for $\hat{G}'$ and $\hat{G}''$ as plotted for $\alpha=1$ 
in Figure \ref{graph:Gs} (a) and for $\alpha=100$ in Figure \ref{graph:Gs}(b). 
Note that both $\hat{G}'$ and $\hat{G}''$ increase as $\omega^{7/8}$ for  $\omega 
\tau_{\xi} \gg 1$. 
 
In addition, 
an intermediate region exists
only if $\tau \gg \tau_{\xi}$ or $\alpha \gg 1$.
From the definition (\ref{eq:alphadef}), this condition is realized in the situation that the total chain length $L$ 
is much larger than $\epsilon^{1/2} \ell_p = \xi^{1/2} \ell_p^{1/2}.$
When this condition is satisfied, 
there is a finite interval of the intermediate region; 
$1/\tau \ll \omega \ll 1/\tau_{\xi}$.
For example, in both 
Figure \ref{graph:Js}(b) and Figure \ref{graph:Gs}(b), 
the interval $1 \ll \omega \tau \lesssim 10^2$ corresponds to this region.
In this region, the asymptotic form of $N$ is given by eq \ref{eq:N0limit}.
Moreover, since $\bar{B}(i \omega \tau_{\xi}) 
= \alpha N(i \omega \tau_{\xi})^{1/2}/2 
\sim (i \omega \tau)^{1/2}$,
the imaginary part of $\bar{B}$ is very large.
Therefore, we can approximate $\tan(\bar{B})$ by $+i$ 
and, substituting eq \ref{eq:N0limit} into eqs \ref{eq:J} and \ref{eq:JJ}, 
the compliance becomes
\begin{gather}
\hat{J}'(\omega) \sim \omega^{-1/2} \frac{f_0^{1/4} (k_BT)^{1/2}}{\sqrt{2}L \zeta_{\parallel}^{1/2} \kappa^{1/4}}
\left( 1-\frac{1}{2} (\omega \tau_{\xi})^{1/2} \right) 
\notag \\
\hat{J}''(\omega) \sim \omega^{-1/2} \frac{f_0^{1/4} 
 (k_BT)^{1/2}}{\sqrt{2}L \zeta_{\parallel}^{1/2} \kappa^{1/4}}
\left( 1+\frac{1}{2} (\omega \tau_{\xi})^{1/2} \right) \ . \label{eq:intermedexp}
\end{gather}
Thus, the compliance has the $\omega^{-1/2}$ dependence
in the intermediate region.

\section{COMPARISON WITH ROUSE DYNAMICS} \label{sec:Rouse}
In this section, following the 
paper
by Khatri and McLeish \cite{Khatri}, 
we present the complex compliance for the Rouse model and compare it with the present result.
The Rouse dynamics without internal friction is governed in the continuum limit by
\begin{equation} \label{eq:Rouse}
\zeta \frac{d {\bm r}(n,t)}{dt} = k \frac{\partial^2{\bm r}(n,t)}{\partial n^2}+{\bm f}(n,t)
+{\bm \xi}(n,t) \ ,
\end{equation}
where 
$\zeta$ is the friction coefficient and
the argument $n$ indicates the $n$-th monomer
from one end, ${\bm r}(n)$ is the position vector of the $n$-th monomer
and $k$ is the elastic coefficient of the linear spring between 
a pair of adjacent two monomers.
It is noted that
the argument $n$ and the number of monomer $N$ are treated as real numbers and
satisfy $0 \leqq n \leqq N$.
Over-damped and Markov motion is assumed.
Both end points are subjected to the external forces which have the same amplitude but the opposite direction
\begin{equation}
 {\bm f}(n,t) =  {\bm f}(t) \left[ \delta(n-N)-\delta(n) \right] \ .
\end{equation}
The last term ${\bm \xi}_n$ in eq \ref{eq:Rouse} is the
White Gaussian noise that satisfies 
the fluctuation dissipation relation of the second kind
\begin{equation} 
  <{\bm \xi}(n,t) {\bm \xi}^{\dagger}(m,t')>=2 k_BT \zeta {\bm I}\delta(n-m) \delta(t-t') \ ,
\end{equation}
where ${\bm I}$ is the unit matrix and two adjacent matrices mean a tensor product.

\begin{figure}[!t]
 \centering
 \includegraphics[width=7.5cm]{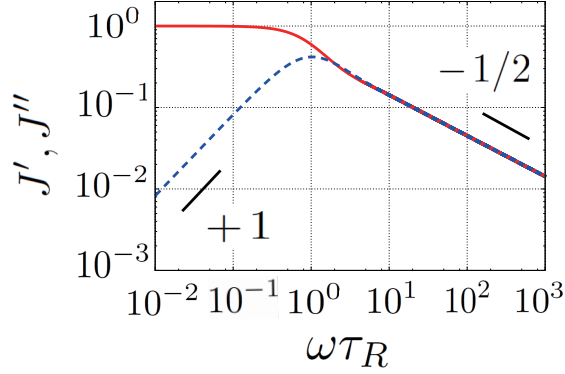}
 \caption{The complex compliance $J_R(\omega)=J_R'(\omega)+i 
 J_R''(\omega)$ for the Rouse dynamics without internal friction.
The full curve represents $J_R'$ whereas the broken curve represents $J_R''$.
The amplitude is scaled 
such that $J_R'=1$
for $\omega \rightarrow 0$.} 
 \label{graph:RouseJ}
\end{figure}

\begin{figure}[!t]
 \centering
 \includegraphics[width=8.0cm]{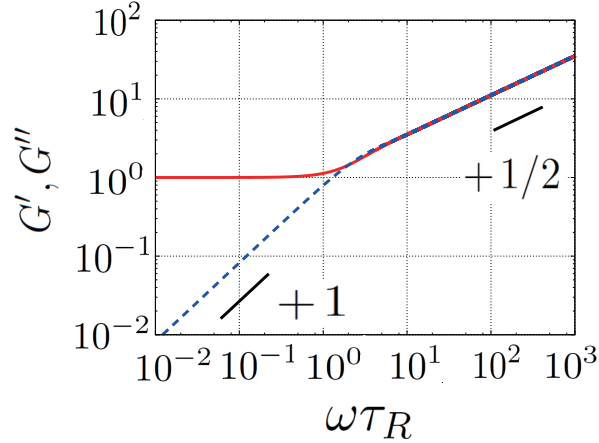}
 \caption{The complex modulus $G_R(\omega)=G_R'(\omega)+i G_R''(\omega)$ 
for the Rouse dynamics without internal friction.
The full curve represents $G_R'$ whereas the broken curve represents $G_R''$. 
The amplitude is scaled such that 
$G_R'=1$ for $\omega \rightarrow 0$.} 
 \label{graph:RouseG}
\end{figure}

We define the end-to-end distance as ${\bm R}(t)={\bm r}(N,t)-{\bm 
r}(0,t)$ and the deviation  as $\Delta {\bm R}(t)={\bm R}(t)-{\bm R}(0)$.
In the same way, the deviation of the external force is defined by $\Delta {\bm f}$.
The complex compliance $J_R(\omega)=J_R'(\omega)+i J_R''(\omega)$ is defined through the relation
\begin{equation}
 <\tilde{\Delta \bm{R}}>(i \omega) = J_R^{*}(\omega) \tilde{\Delta {\bm 
  f}}(i \omega) \ ,
\end{equation}
where the asterisk $\ast$ 
means the complex conjugate and $J_R^{*}(\omega)$ is
given by \cite{Khatri}
\begin{equation}
 J_R^{*}(\omega) = \frac{2N}{\pi k} \frac{\tanh\left(\frac{\pi}{2} \sqrt{i 
 \omega \tau_R} \right)}  {\sqrt{i \omega \tau_R}} \ ,
\label{eq:RouseResult}
\end{equation}
where $\tau_R$ is the Rouse relaxation time 
defined by
\begin{equation}
 \tau _R = \frac{N^2 \zeta}{\pi^2 k} \ .
\end{equation}
The function  (\ref{eq:RouseResult}) is plotted in Figure \ref{graph:RouseJ} and
the corresponding complex modulus $G_R$ is plotted in Figure \ref{graph:RouseG}.

From the expression (\ref{eq:RouseResult}), the asymptotic behavior is derived  to compare with that of the weak-bending wormlike-chain dynamics.
For $\omega \rightarrow \infty$, the complex compliance behaves as
\begin{align}
 J_R'(\omega) &\propto \omega^{-1/2} \\
 J_R''(\omega)&\propto \omega^{-1/2} \ ,
\end{align}
and for $\omega \rightarrow 0$ 
\begin{align}
 J_R'(\omega) &\rightarrow \text{const.} \\
 J_R''(\omega)&\propto \omega^{+1} \ .
\end{align}
These exponents 
are distinctly different from these
obtained in the previous section,
 $-7/8$ in both $J_R'$ and $J_R''$ as $\omega 
\rightarrow \infty$ and $+1/2$ in $J''$ as $\omega \rightarrow 0$.
See eqs \ref{eq:winftyexp} and \ref{eq:NbarS2}.
Moreover, the viscoelastic behavior of the Rouse dynamics with internal friction is 
also examined by Khatri and McLeish \cite{Khatri},
where the high frequency behavior is given by $J_R' \propto \omega^{-2}$ and $J_R'' \propto \omega^{-1}$. 
These are again different from the present results.

\section{SCALING APPROACH} \label{sec:scaling}
\subsection{Scaling form of $\Delta R(t)$}
In this section, we apply the scaling analysis in order to 
explain the behavior of complex compliances for both high and low frequency limits.

All the parameters are scaled out in 
eqs \ref{eq:nonlinear5} and \ref{eq:eliofeps}. 
Therefore, the parameters appear only through eq \ref{eq:FhatFtrans} and 
through the boundary condition which contains $L$.
The scaled form of $L$ is given by
 $\hat{L} =\epsilon^{1/2}L \xi^{-1}$. These facts together with eq \ref{eq:DelR} 
give us the following scaling property of $\Delta R$;
\begin{equation} \label{eq:DRscform}
 \Delta R(t) = \epsilon^{1/2}\xi^{-1} \tau_{\xi} f \zeta^{-1} 
  Q(\epsilon^{1/2}L\xi^{-1}, t/\tau_{\xi}) \ ,
\end{equation}
where $Q(x,y)$ is an unknown function whose asymptotic form is to be determined.
This scaling form (\ref{eq:DRscform})
is very crucial to investigate the asymptotic behavior of the 
compliance and the modulus as shown below.

\subsection{Complex compliance for the high frequency limit}
The exponent $7/8$ exhibited by $J'$ and $J''$ for $\omega \rightarrow \infty$ obtained 
in eq \ref{eq:winftyexp} can be understood by the following scaling analysis.
In the linear response regime, the dimensionless function $Q$ in eq \ref{eq:DRscform} 
should be proportional to $f_A/f_0$. 
At the high frequency limit, 
the effect of the external force
is expected to be localized near the two ends
and hence the compliance $J$ should not depend on $L$. 
Therefore, from eq \ref{eq:DRscform}, the compliance takes the following form
\begin{equation} \label{eq:highwscaling}
 f_A J(\omega) \sim \frac{f_A}{f} \epsilon^{1/2}\xi^{-1}\tau_{\xi} f 
  \zeta^{-1} (\omega \tau_{\xi})^{-z}
\end{equation}
with an unknown exponent $z$.
Substituting the definitions of 
$\xi$, $\epsilon$ and  $\tau_{\xi}$ given, respectively, by (\ref{eq:xi}), 
(\ref{eq:epsreform}) and (\ref{eq:tauN}) into eq \ref{eq:highwscaling} yields
\begin{equation} \label{eq:scalingJhighfreq}
 f_A J(\omega) \sim f_A
f^{-7/4+2z} \kappa^{1/4-z} (k_BT)^{1/2} 
  \zeta^{-z} \omega^{-z} \ .
\end{equation}
We can require that the compliance is independent of the screening length 
$\xi$ (and hence $f$) in the high frequency limit.
This is because the relaxation of the chain has the factor $\kappa q^4+f q^2$ 
as can be seen from eq \ref{eq:dynamics1} or eqs \ref{eq:defA} and \ref{eq:Fst}
and hence $\kappa$ is relevant for the high frequency ($fq^2$ is irrelevant).
This requirement gives us
\begin{equation}\label{eq:z}
 z = \frac{7}{8} \ .
\end{equation}

The scaling analysis in the Rouse dynamics is 
different from the above because of the absence 
of the local length scale, i.e., the persistence length 
$\ell_p$.
The Rouse model has only one length scale, the root of the mean square end-to-end 
distance $\sigma$.
When no external force is present, it is given by \cite{DoiEd}
\begin{equation}
 \sigma \sim \left( \frac{k_BT}{k}\right)^{1/2} N^{1/2} \ .
\end{equation}
Therefore the dimensional analysis tells us that the deviation of the end-to-end distance should 
obey
\begin{equation}
 \Delta R \sim \sigma \frac{\sigma f_A}{k_BT} \hat{J}(\omega \tau_R) e^{i \omega t}
\end{equation}
with the external force
\begin{equation}
 f(t) = f_A e^{i \omega t} \ .
\end{equation}
Assuming that $\hat{J}$ has a power law behavior
$\hat{J}(\omega \tau_R) \sim (\omega \tau_R)^{-z}$ as $\omega \rightarrow \infty$.
The complex compliance becomes
\begin{equation}
 f_A J(\omega) \sim \sigma \frac{\sigma f_A}{k_BT}(\omega \tau_R)^{-z} \sim 
  f_A N^{1-2z} \omega^{-z} k^{z-1}\zeta^{-z} \ .
\end{equation}
In the high frequency limit, the response is localized and 
the compliance
should be independent of $N$ so that the exponent is determined uniquely as $z=1/2$ or
\begin{equation}
 J(\omega) \propto \omega^{-1/2} \ .
\end{equation}
This is the argument given  by Khatri et al. \cite{Khatri}.

\subsection{Complex compliance for the low frequency limit}
The exponent $1/2$ exhibited by $J''$ for $\omega \rightarrow 0$ obtained in
eq \ref{eq:NbarS2} can be understood as follows.
In the low frequency limit, the effect of the external force is extended almost uniformly to the whole chain. 
Therefore, we can require that $J$ is proportional to $L$ so that
\begin{equation} \label{eq:lowwscaling}
f_A J(\omega)=  \frac{f_A}{f} \epsilon^{1/2}\xi^{-1}\epsilon^{1/2}L\xi^{-1} 
 \tau_{\xi} f \zeta^{-1}(\omega \tau_{\xi})^z \ .
\end{equation}
The real part $J'$ is independent of the frequency for $\omega \to 0$ and 
hence $z=0$ whereas
the imaginary part $J''$ should be independent of $\kappa$ for $\omega \to 0$. 
As mentioned above, the relaxation of the chain has the factor $\kappa q^4+f q^2$ and hence
$\kappa$ is irrelevant for the low frequency.
Therefore, substituting the definitions of 
$\xi$, $\epsilon$ and  $\tau_{\xi}$ given, respectively, by (\ref{eq:xi}), (\ref{eq:epsreform}) and (\ref{eq:tauN})
into eq \ref{eq:lowwscaling},
it is found that the exponent is given by
\begin{equation}\label{eq:z2}
 z = \frac{1}{2}
\end{equation}
so that 
$J'' \propto k_BT \omega^{1/2}$.
Note that, from eq \ref{eq:scalingJhighfreq}, $J''$ at the high frequency limit is proportional to $(k_BT)^{1/2}$ 
whereas it is proportional to $k_BT$ at the low frequency limit. 

In contrast,  the complex compliance in  the Rouse 
dynamics is analytic
in the $\omega \rightarrow 0$ limit.
This fact is clear from the expression 
(\ref{eq:RouseResult}).
This should be compared with that of the wormlike-chain dynamics
(\ref{eq:NbarS1}) and (\ref{eq:NbarS2})
where the function 
N(z) in $\bar{B}(z)$ contains a non-analyticity as eq \ref{eq:N0limit}.

\section{SUMMARY AND DISCUSSION} \label{sec:con}
In summary, we have developed the analytical theory of the viscoelasticity  
of single semiflexible polymer chains and have  obtained the linear compliance which has a frequency dependence characteristic to the semiflexible chain.  
In particular, it is found that the asymptotic behavior of the compliance obeys 
as $J', J'' \propto \omega^{-7/8}$ for $\omega \to \infty$ 
whereas $J'' \propto \omega^{+1/2}$ for $\omega \tau \ll 1$. 
These are distinctly different from the results of the Rouse dynamics. 
The constant of $J_R'$ for $\omega =0$ given by eq \ref{eq:NbarS1}
 is also different from that of the flexible chain.

The theory assumes  weakness of the  bending parameter $\epsilon \ll 1$ which guarantees the scale separation.
This is due to the fact that the characteristic length parallel to the stretched chain $\ell_{\parallel}
\sim \Delta s$ and the 
characteristic wave length $q^{-1} \sim \ell_{\perp}$ satisfy
\begin{equation}
 \frac{\ell_{\perp}}{\ell_{\parallel}} \sim \frac{q^{-1}}{\Delta s} \sim 
\epsilon^{1/2} \frac{\hat{q}^{-1}}{\Delta \hat{s}} \ll 1 \ ,
\end{equation}
where the scaling forms (\ref{eq:qsscale}) and (\ref{eq:eliofeps})
have been used.
It is emphasized
that, for $\omega \to \infty$, the scale separation is valid without 
assuming the smallness of $\epsilon$ because of the fact that $\ell_{\parallel}
\propto \omega^{-1/8} \gg \ell_{\perp} \propto \omega^{-1/4}$.
This fact is verified by eqs \ref{eq:raplace} and \ref{eq:Nbar}.

Now we discuss the relation between the present results and those obtained by
Hallatschek et al. who have considered the relaxation of the end-to-end distance
 after step-wise change of the external force \cite{Hall2,Ober2}. 
They have predicted that both in the stretching case and in the release
case the end-to-end distance behaves as
\begin{equation}\label{eq:Hall}
<\Delta R_{\parallel}(t)> \propto f 
\kappa^{-5/8} (k_BT)^{-1/2} t^{7/8} \ ,
\end{equation}
where $t \ll t_f \sim \zeta \kappa f^{-2}$.
The exponent $7/8$ is the same as that in the high-frequency limit 
in eq \ref{eq:winftyexp}.
At a short interval after the force change, its effect 
is localized near the chain ends which is small compared with
both the persistence length (\ref{eq:lp})
and the screening length (\ref{eq:xi}).
In fact, we can show that eq \ref{eq:Hall} is consistent 
with our eq \ref{eq:winftyexp} as follows.
In the linear response theory
the relaxation function $\psi(t)$ and response function $\phi(t)$
are related to
each other as $\phi(t) = - d\psi(t)/dt$.
The complex compliance is the Fourier-Laplace transform of
the response function $\phi$ and hence $J(\omega) = \psi(0) + i\omega
\int_{0}^{+\infty} dt e^{i\omega t} \psi(t)$.
Therefore, we have the following relation $J(\omega) \sim \psi(1/\omega)$
and $\hat{J}(\omega) = f J/L \sim \langle \Delta R_{\parallel} 
(1/\omega)\rangle /L$.

In the intermediate time region $t_L \gg t \gg t_f$ 
with the crossover time $t_L$ defined 
through $\ell_{\parallel}(t_L) = L$, 
Hallatschek et al. have 
obtained \cite{Hall2}
\begin{equation} \label{eq:pull}
<\Delta R_{\parallel}(t)> \propto f^{3/4}
\kappa^{-1/2} (k_BT)^{1/2} t^{3/4}
\end{equation}
for a pulling situation and
\begin{equation}\label{eq:rele}
<\Delta R_{\parallel}(t)> \propto f^{1/4} 
\kappa^{-1/4} (k_BT)^{1/2} t^{1/2}
\end{equation}
for a release situation.
We have no results  corresponding to eq \ref{eq:pull} since this contains a  nonlinear effect of the 
applied force. On the other hand, the exponent 1/2 in eq  \ref{eq:rele} corresponds  
to eq \ref{eq:intermedexp}
in the present paper. Actually one can verify that not only the exponent but also  
the coefficient in eq \ref{eq:rele} is consistent with our result. 
This implies that the expression (\ref{eq:rele}) is free from the nonlinearity between the force-strain relation.

Finally we mention a theoretical study which 
gives us the exponent 1/2 in the compliance.
Caspi et al. have investigated the mean square displacement of a single monomer
of a prestressed semiflexible network \cite{Caspi}.
They have obtained 
\begin{equation}\label{eq:Caspi1}
 <\Delta h^2(x,t)> \propto \frac{k_B T}{\nu^{1/2} \eta^{1/2}} t^{1/2} \ ,
\end{equation}
where $h(x,t)$ denotes the undulation amplitude, $\nu$
the line tension, $\eta$ the solvent viscosity and $L$
the total chain length. Equation (\ref{eq:Caspi1}) holds in the time region $4\pi\eta\kappa/\nu^2 \ll t \ll \eta L^2/\nu$.
Furthermore, they have shown that the effective time dependent friction 
$\zeta_e(t) $ satisfies the generalized Einstein relation
\begin{equation} \label{eq:Caspi2}
 \frac{k_B T}{\zeta_e(t)}=\frac{<\Delta h^2(t)>}{2t} \ .
\end{equation}
Combining eqs \ref{eq:Caspi1} and \ref{eq:Caspi2}, one obtains the complex compliance ($J\propto t/\zeta_e(t)$)
\begin{equation}\label{eq:Caspi3}
 J(\omega) \propto \nu^{-1/2} \eta^{-1/2} \omega^{-1/2} \ . 
\end{equation}
Some experiments of semiflexible networks support the exponent 1/2 \cite{Caspi,Mizuno}.
It is mentioned, however, that the physical
of this result is different from our present result for semi-flexible chain 
given by eq \ref{eq:intermedexp}.
This fact is clear because the coefficient in eq \ref{eq:Caspi3} does not contain $k_BT$ whereas
our expression eq \ref{eq:intermedexp} is proportional to  $(k_BT)^{1/2}$.

Now we comment on 
the several effects which have
not been considered in the present paper.
The hydrodynamic effect 
has not been investigated quantitatively in a nonlinear wormlike-chain
although it is expected to be not so strong in a strongly stretched semi-flexible chain.
The previous studies of the hydrodynamic effect 
in the linearized wormlike-chain dynamics \cite{Winkler1,Harnau,Winkler2,Winkler3} should be extended to apply to the present theory.
The internal friction considered in the Rouse dynamics \cite{Khatri} should also be extended to the semi-flexible chains.
In addition, the helical wormlike-chain model, which contains the torsional energy,
has been studied in dilute solutions
\cite{Chirikjian,Yamakawa1,Yamakawa2}.
This torsional effect may affect the viscoelastic properties of single polymer chains.

Before closing this article, we make an estimation of the characteristic times 
$\tau_{\xi}$ and $\tau$ defined by eqs \ref{eq:tauN} and \ref{eq:tau} respectively.
The data for $\lambda$-DNA in an aqueous solution are as follows  \cite{Quake, Maier, Hall2}
\begin{align}
 &\ell_P \sim 50 [\text{nm}], \notag \\
 &L \sim 20 [\text{$\mu$m}], \notag \\
 &\zeta _{\perp} \sim 1.3 \times 10^{-3} [\text{Pa s}] = 1.3 \times 
 10^{-3} [\text{pN$\cdot$s/$\mu$m$^2$}] \ . \notag
\end{align}
For these values together with $\hat{\zeta} \sim 1/2$ for  a rigid rod \cite{DoiEd} 
and the room temperature $ k_B T \sim 4.1 [\text{pN}\cdot\text{nm}] $, and for the external force $f_0 \sim 10 [\text{pN}]$
the characteristic times are given by
\begin{align}
 &\tau_{\xi} \sim 2.7 \times 10^{-9} [\text{s}] \notag \\
 &\tau \sim 6.0 \times 10^{-5} [\text{s}] = 60 [\text{$\mu$s}] \notag
\end{align}
and the constant $\alpha \sim 5.3 \times 10^2 $. 
We expect that the frequency of the order of 60 [\text{$\mu$s}] is accessible by 
atomic force microscopy and that the present predictions can be detected experimentally.

\section*{acknowledgments}
This work was supported by 
the Grant-in-Aid for priority area "Soft Matter Physics" 
from the Ministry of Education, Culture, Sports, Science and Technology (MEXT) of Japan.
The scaling theory was completed during TO's stay in Institut 
f\"{u}r Festk\"{o}rperforschung, J\"{u}lich and in University of Bayreuth. 
The financial support from the Alexander von Humboldt foundation is gratefully acknowledged.

\end{document}